\documentstyle[preprint,prb,aps,
epsfig]{revtex}
\begin{document}
\headheight2.0cm
\headsep1.2cm
\baselineskip8mm
\tolerance=1500
\thispagestyle{empty}
\begin{list}{}{\setlength{\leftmargin}{1mm}\baselineskip8mm}
\vglue.01cm
\item
\begin{center}
\item LOW ENERGY EXCITATIONS OF DOUBLE QUANTUM DOTS IN THE
\item LOWEST LANDAU LEVEL REGIME.
\vskip3mm
\item N. Barber\'an and J. Soto
\item Departament d'Estructura i Constituents de la Mat\`eria
\item Facultat de F\'\i sica, Universitat de Barcelona,
E-08028 Barcelona, Catalonia, Spain
\end{center}
\vskip7mm
\begin{abstract}

We study the spectrum and magnetic properties
of double quantum dots in the lowest Landau level for different values of the
hopping and Zeeman parameters by means of exact diagonalization
techniques in systems of N=6 and N=7 electrons and filling factor
close to 2. We compare our results
 with those obtained in double quantum layers and single quantum dots.
The Kohn theorem is also discussed.
\end{abstract}

\vskip10mm

KEYWORDS:  Quantum Hall effect, double quantum dot, multipolar
excitations.

PACS:  73.21-b, 73.43-f, 73.21.La, 73.43-Lp.
\end{list}
\vfill
\vbox{
\hfill April 2002\null\par
\hfill UB-ECM-PF 02/10}\null\par

\eject

\subsection*{I. Introduction}

The understanding of the structure and properties of double quantum
dots (DQD's) grown on the perpendicular direction to their plain, and
subject to an external constant magnetic field $\vec{B}$, has
attracted
special interest due to the wealth of new quantum states that
are actually realized. This wealth
does not only refer to the ground states (GS's), which
provide a quite intricate phase diagram \cite{mar,par}, but also
to the variety of different excited states.
This allows to have a
large quantity of different sets of properties of the DQD system that
can model for instance, point contacts within a device built for
electron transport.

\medskip

Special interest has the regime for which the magnetic
field is so strong that only the lowest Landau level (LLL) is occupied
but
not strong enough that prevent any spin polarization. We will always
assume
that the second Landau level is far enough that we can ignore any
mixture between Landau levels.
In the symmetric gauge, the projection of the DQD Hamiltonian to the
LLL is given by \cite{tej}

\begin{equation}
H=\alpha M + \beta N - \Delta_z S_z - \Delta_t X +
H_{int}
\label{ham}
\end{equation}
where

\begin{equation}
\alpha = \frac{\hbar}{2}\left[ \sqrt{\omega_c^2 + 4 \omega_0^2}-
\omega_c\right]
\label{alf}
\end{equation}
and

\begin{equation}
\beta=\frac{\hbar}{2} \sqrt{\omega_c^2+4 \omega_0^2}\,\,\,,
\label{bet}
\end{equation}
$\omega_c=\frac{eB}{cm^*}\,\,$ is the cyclotron frequency, $m^*$ being
the
effective electron mass in the semiconductor host and $e$ and $c$
are the electron charge and
the speed of light in vacuum respectively. The
frequency
associated with the parabolic confining potential in both
dots is given by $\omega_0$,
$M$
is the total angular momentum and $N$ is the total number of
electrons.
The Zeeman coupling is given by
$\Delta_z=g\mu_B B$  with $g$
being the Land\'e $g$-factor and $\mu_B$ being the Bohr magneton ( for
the free electron mass, $\mu_B=e\hbar/2mc$).
The
single particle energy gap between symmetric (s) and antisymmetric
(a)
states, combinations of $\mid$$r$$\rangle$ and $\mid$$l$$\rangle$
single particle
states of electrons confined in the right and left dot respectively
(see
below) is given by
$\,2\Delta_t\,\,$  and
$X=N_s-N_a$ is the balance between symmetric and
antisymmetric single particle states. Finally,
$H_{int}$ is the Coulomb interaction term.
Hereafter all distances
will be given in units of the magnetic length defined as:

\begin{equation}
l_B=\left[ \frac{\hbar}{m^*( \omega_c^2+4
\omega_0^2)^{1/2}}\right]^{1/2}
\label{lb}
\end{equation}
and all  energies in units of $e^2/(\epsilon l_B)$ being $\epsilon$
the dielectric constant of the semiconductor host.
The Zeeman energy, the tunneling
term scaled by $\Delta_t$, the kinetic contribution given by
$\alpha$ and $\beta$ and the Coulomb interaction, provide the
ingredients of the system.
The applied magnetic field can be directed in any direction in space
in such a way that its action has different effect in the kinetic
contribution in which only the component along the $Z$-direction plays
a role and in the Zeeman effect in which the total magnetic field
contributes.
The eigenstates of the Hamiltonian are
characterized by the total angular
momentum $M$, the total spin along the $\vec{B}$-direction given by
$S_z$
and the parity $P$ \cite{mar,oak} . These parameters are related with
the invariance of the Hamiltonian
under space rotations along the $Z$-direction, rotations of spin and
specular reflection in the plane inter-dots respectively \cite{mar}. We
will denote by $(M,S_z,P)$ the configuration that
determines each separated subspace of eigenstates.

\medskip

Previous studies in DQD's based on exact diagonalizations on
one hand
\cite{mar,par} and the large experience extracted from mean field
(MF) approximations and effective field theories
in double layer (DL) systems on the other hand
\cite{moo,gir,das,son,par1},
have provided a quite complete picture of the GS phase diagram.
It is well established in both DL (Ref. 7) and DQD (Ref. 1)
systems that the variation of
$\,\Delta_t\,\,$, leaving
fixed all the other parameters, induces several changes in the GS.
For
$\,\Delta_z\gg \Delta_t\,$ the GS is ferromagnetic $\mid$FM$\rangle$
with
symmetric and antisymmetric single particle states equally populated
and with all the electrons in spin up states. In contrast for
$\,\Delta_z\ll\Delta_t\,$ the GS is symmetric $\,\mid$SYM$\rangle\,$
with all the
electrons occupying symmetric single particle states with up and down
spins equally populated. In between, for comparable values of
Zeeman
and tunneling energies, intermediate states of the type
called canted states ($\mid$C$\rangle$) become GS's.
The canted state is the GS solution between the $\mid$FM$\rangle$
and
$\mid$SYM$\rangle$
in a MF calculation for a DL system and it is characterized by a
ferromagnetic order in the direction perpendicular to the layers and
antiferromagnetic order in plane \cite{das}.

\medskip

A remarkable result of the Hartree Fock (HF) approximation for DL
quantum Hall systems at filling factor $\nu=2$ is that there are
only three possible different GS's: $\mid$FM$\rangle$,
$\mid$C$\rangle$ and $\mid$SYM$\rangle$ distributed over a universal
phase diagram which boundaries depend only on three energy scales
whereas, the Hamiltonian depends on four independent energy scales
\cite{das}.

\medskip

Certain ground states obtained from exact diagonalization in DQD's
have
also been identified as canted states since their overlap with the MF
canted GS's for finite systems,
projected on the appropriate subspace with
well defined $S_z$,
is nearly one \cite{mar}.
The total number of different canted states between $\mid$FM$\rangle$
and
$\mid$SYM$\rangle$ in a DQD
depends on the number of electrons. As a general rule, keeping fixed
$\Delta_z$ and increasing $\Delta_t$, the GS evolves as:

\begin{equation}
\mid FM\rangle \longrightarrow (M, S_z,P) \longrightarrow (M,S_z-1,
-P)
\longrightarrow
(M,S_z-2
,P),\,...\,,
\mid SYM\rangle
\label{fm}
\end{equation}
A simultaneous change of spin and parity occurs for the GS
in each phase
transition whereas the total angular momentum remains unchanged
\cite{mar}.

\medskip

No direct information about the order in plane has been obtained
so far from
the exact solutions due to the fact that they have all the quantum
numbers ($M, S_z,P$) well defined. No information can directly be
obtained from order parameters that depend on single operators of the
type $S_x^R$ or $S_x^L$ since their expectation values vanish.
However, indirect information about the order in plane can be obtained
from the properties of the excited states as we will show in the
discussion of Figs.(1) and (2) below.
Within the great number of possible low energy excitations that
provide
information about the properties of the system, we will draw our
attention to those related to three main different points:

\medskip

First,
we search
for particle-like excitations
independent of the electron-electron interaction
and so independent of the number of electrons. We discuss when the
Kohn theorem \cite{koh} holds in a parabolic DQD system ( as it is
always the case for a single parabolic QD)

\medskip

Second,
we look for the types of excitations that soften as
they come close to a GS transition that takes place with the variation
of some input parameter. These excitations provide a clear and easy
way to map the GS phase diagram.

\medskip

Third, in the last point
we concentrate in
the dispersion relation $\omega (l)$ of two different type of
excitations, one with and the other without a simultaneous spin and
isospin flip (see below). In a single QD the Coulomb contribution to
the dispersion relation of excitations over a ferromagnetic GS
decreases with angular
momentum due to expansion. This is the general behavior, except at
the "magic" values of $l^*$ \cite{jac} for which the system
increases in angular momentum from $l^*$ to $l^*+1$ but does not
increase in size and hence the Coulomb contribution remains constant.
These magic
values of $l$ are related to the incompressible states. Our aim is to
see if there is a similar behavior in the DQD system.

\medskip

This paper is organized as follows: In Section II we
make a brief account of the exact diagonalization  used in
our calculations. In Section III we present the outstanding
features of some
excited states of different multipolarity over different GS's, paying
special attention to the different behavior of even and odd systems.
We will follow the three main points previously mentioned.
Finally in Section IV we draw our
conclusions.

\subsection*{II. Exact diagonalization}

Exact diagonalization can be performed in separate boxes
for different configurations characterized by $(M,S_z,P)$. Each
configuration determines a finite-eigenstate subspace.
Within the LLL regime, we will use in all expansions
the non-interacting
single particle wave functions which do not have
nodes along the radius and are given by

\begin{equation}
\Phi_m(\vec{r},\sigma,\Lambda)\,\,=\,\,\phi_{m}(\vec{r})\,\,\mid
\sigma
\Lambda\rangle
\label{phi}
\end{equation}
where $m$ is the single particle angular momentum,
$\,\Lambda\,=\,r,l,\,\,$
$\sigma=$ $\uparrow$, $\downarrow$, $\,\, \vec{r}=(x,y)$,
and $\,\phi_m(\vec{r})\,$ are the Fock Darwin wave functions
given by \cite{jac}

\begin{equation}
\phi_{m}(\vec{r})=\frac{1}{\sqrt{2\pi m!2^m}} e^{-im\theta}r^m
e^{-r^2/4}\,\,\,\,.
\label{phr}
\end{equation}
Along the $Z$-direction we assume delta charge distributions
separated by $d$, the distance between the dots (we
consider
$d=1$ in all numerical performances).

\medskip

Within our calculations we will use symmetric $\mid$$s$$\rangle$ and
antisymmetric
$\mid$$a$$\rangle$ single particle states, related by
$\mid$$r$$\rangle$ and
$\mid$$l$$\rangle$ by

\begin{equation}
\mid s\rangle\,\,=\,\,\frac{1}{\sqrt{2}}(\,\,\mid r\rangle
\,\,+\,\,\mid l\rangle\,\,)
\label{s}
\end{equation}
and

\begin{equation}
\mid a\rangle\,\,=\,\,\frac{1}{\sqrt{2}}(\,\,\mid r\rangle
\,\,-\,\,\mid l\rangle\,\,),
\label{a}
\end{equation}
we will use the term isospin referred to this degree
of freedom (sometimes referred as pseudospin in literature,
see Ref.\cite{moo}).
None of the parameters $r$ and $l$ or $s$ and $a$ are well defined
quantum numbers since the Coulomb interaction mixes $s$ and $a$
and the tunneling process mixes $r$ and $l$. The well established
restriction is that parity must be preserved, this means that the
change in symmetry ( $s\rightarrow a$ or $a\rightarrow s$) due
to electron interaction
must take place by pairs of electrons, and never by one alone (see
below).

\medskip

We will consider excited states over the three possible types of GS's
and choose the parameters in such a way that for an even number of
electrons in the ferromagnetic GS we have filling factor $\nu=2$.
We proceed as follows. Once the input parameters are fixed, we
determine first the finite number of Slater determinants built up from
single particle wave functions of the type given by Eq.(\ref{phi})
which provide a bases for each subspace configuration $(M,S_z,P)$. Then,
the diagonalization of the Hamiltonian given by Eq.(\ref{ham}) is
straightforward except for the Coulomb term. Although the Coulomb
interaction
does not mix $\mid$$r$$\rangle$ and $\mid$$l$$\rangle$ single particle
states, some manipulations must be done in order to operate
over $\mid$$s$$\rangle$ and $\mid$$a$$\rangle$ wave functions.
In second quantization, the interaction Hamiltonian is given by

\begin{equation}
H_{int}=\frac{1}{2} \sum_{ijkl} \langle \phi_i \phi_j\mid V \mid
\phi_k \phi_l\rangle a^+_i a^+_j a_l a_k
\end{equation}
where $V$ is the Coulomb interaction and the sub-indexes denote angular
momentum $m=0,1,2,..$, spin=
$\uparrow,\downarrow$ and isospin $\tau$= $a$, $s$. Taking into
account
that the single particle wave functions are related by Eq.(\ref{s})
and (\ref{a}) and that the right and left single states are
$\delta$-distributions of the type $\delta(z)$ and $\delta(z-d)$
respectively, it is easy to show that

\begin{equation}
\phi_a^*\,\phi_a\,=\,\phi_s^*\phi_s\,\sim\,\frac{1}{2}(\delta(z) \,
+\delta(z-d))
\end{equation}
and

\begin{equation}
\phi_s^*\,\phi_a\,\sim\,\frac{1}{2}(\delta(z)-\delta(z-d))
\end{equation}
for each electron. As a  consequence, there are only three possibilities
for the expectation values of $V$: (i) the interacting electrons do not
change their isospin as in
\begin{equation}
\langle \phi_s\phi_s\mid\,V\,\mid \phi_s\phi_s\rangle =
\frac{1}{2}(\,\langle V_{rr}\rangle\,+\,\langle V_{rl} \rangle)
\end{equation}
(ii) only one electron changes its isospin as in,
\begin{equation}
\langle\,\phi_s\phi_a\mid V\mid \phi_a\phi_a\rangle = 0
\end{equation}
and (iii) both electrons change their isospin index as in
\begin{equation}
\langle \phi_s\phi_a\,\mid\,V\,\mid \phi_a\phi_s\rangle =
\frac{1}{2}(\langle V_{rr}\rangle-\langle V_{rl}\rangle)\,\,\,.
\end{equation}
In the brackets on the right hand side of these equations, the
integral over the $Z$-coordinate has been performed and the
potentials
are given by
\begin{equation}
V_{rr}=V_{ll}=\frac{e^2}{\epsilon
r},\,\,\,\,\,\,\,\,\,\, \,\,\,\,\,\,\,\,\,V_{rl}=V_{lr}
=
\frac{e^2}{\epsilon (r^2+d^2)^{1/2}}
\end{equation}
with $r=\mid\vec{r}_1-\vec{r}_2\mid$, 1 and 2 being the two
interacting electrons.
That is to say, the Coulomb interaction either leaves
unchanged the
 isospin of the electrons, if it operates with
\begin{equation}
V_0=\frac{1}{2}(V_{rr}\,+\,V_{rl})
\end{equation}
or changes the isospin of two electrons, if it operates with
\begin{equation}
V_1=\frac{1}{2}(V_{rr}\,-\,V_{rl})
\end{equation}
in such a way that parity $P$, given by $P=(-1)^{X/2}$
is preserved ($X=N_s-N_a$) \cite{mar} . The change of
isospin of a single electron is forbidden.

\subsection*{III  Multipolar excitations}

Guided by an interest in the properties of the multipole mode
excitations and also by an interest in those excitations that soften
at the boundaries of the phase transitions within the phase diagram,
we explored the eigenenergies and eigenstates
coming out of exact diagonalizations.
We consider, in general, the multipolar excitations
characterized by $\omega (l)$ with or without a simultaneous change
of spin and/or parity and in some cases we concentrate on $l=0$ or
$l=1$.

\medskip

Before focusing on the first point,
that is,
within the search
of excitations which keep the internal Coulomb energy
constant,
let us briefly comment on the Kohn theorem for a DQD.
The Kohn theorem for a single QD states \cite{bre} that in
the process of absorption (or
emission)
of a photon of wavelength much larger than the radius of the dot
confined by a parabolic potential, the initial and final electronic
states can differ from one another only by the center of mass (CM)
excitation. The
number of confined electrons and the interaction between them has no
influence on the values of resonance energies.
For a DQD, however,
qualifications to this statement may be required
for different
directions of the incident electric field
with respect to the plane of the dots. It is easy to see that
for homogeneous electric fields in-plane the
Kohn theorem also holds for DQD's.
However, this is not the case for homogeneous electric fields  along
the $Z$-direction, as we argue below.

\medskip

Taking into account Eqs.
(\ref{s})-(\ref{a}) and the expression of the exciting potential
(associated with $E_{\perp}$ directed along the $Z$-direction)

\begin{equation}
H_{E_{\perp}}\sim \sum_{i=1}^N z_iE_{\perp}\,\,\,\,\,,
\end{equation}
it is easy to see that $\langle \phi_s \mid H_{E_{\perp}} \mid \phi_a \rangle \sim
d\neq
0$.
That is to say, the operator $H_{E_{\perp}}$ changes the parity of the
system and
so $[\,H_{E_{\perp}},\,H_{int}]\neq 0$ since the eigenstates of
$H_{int}$
have well
defined parity. The part of the Hamiltonian which depends on the
relative coordinates only receives
other contributions aside from the interaction Hamiltonian
(coming from the kinetic and from the tunneling terms
\cite{jac} within the first quantized expression)
however the possibility that the last bracket can be compensated by
the brackets of the other terms can be disregarded as they depend on
different independent parameters. An alternative reasoning is presented
in Appendix A.
In
the analysis of the
exact diagonalization results which follows, we will verify,
among other things,
the applicability of the Kohn theorem.

\medskip

Let us begin with the N=7 electron system and consider first the
excitations
of the $\mid$FM$\rangle$  GS that increase the angular momentum
in one unit and leave all the other parameters of the configuration
unchanged, that is to say, we consider,

\begin{equation}
(M,S_z,P)\,\longrightarrow \,(M+1,S_z,P).
\end{equation}
The system jumps from a 1-dim space configuration to a 2-dim space.
The result is the excitation of the CM by $\alpha$
leaving the
internal Coulomb energy unchanged. This is the well known intra-Landau
level dipole excitation whose energy decreases when the
magnetic field increases that is, the $\,\omega_{-}=\alpha\,\,$
far-infrared resonance (FIR)
\cite{jac}. No inter-Landau level transition of energy given by

\begin{equation}
\omega_{+} = \frac{1}{2}\left[ \sqrt{\omega_c^2 + 4 \omega_0^2}+
\omega_c\right]
\end{equation}
is possible within the LLL considered in our calculation.
Other Coulomb-independent interaction excitations are
possible in the N=7 system. If we also allow changes in parity,
that is to say, if we consider,

\begin{equation}
(M,S_z,P)\,\longrightarrow \,(M+1,S_z,-P)
\end{equation}
the CM is excited with an energy $\alpha\,+\,\Delta_t$ (it
is a CM excitation since the same result would be obtained for N=1).
The
previous
case (of energy $\alpha$) would correspond to the excitation made by
a nearly constant electric field ( we are assuming dipole
approximation) directed along the $X$ direction ($E_{\parallel}$)
and the
last one (of energy $\alpha\, +\, \Delta_t$) would correspond to an
electric field with an additional non-vanishing component
along the $Z$ direction (let us call it {\bf $E_{xz}$}).
Incidentally, this result can be used to determine
experimentally the relative orientation of the DQD respect to the
incident beam. The system absorbs a photon of energy
$\alpha + \Delta_t$ with maximum probability when the angle between
the direction
of the incident electric field and the normal to the plane of the DQD
is $\theta=\pi/4$.

\medskip

For both excitations, with and without parity change, the system goes
from a 1-dim space ($\mid$FM$\rangle$) to a 2-dim space of excited
states.
The CM excitation leaves the system in the higher energy
state
within the excited
2-dim configuration in such a way that,

\begin{equation}
E_2(M+1,S_z,\pm
P)\,\,=\,\,E_1(M,S_z,P)\,\,+\,\,\alpha
\,\,+\,\,(1\mp 1)\,\Delta_t/2
\end{equation}
and

\begin{equation}
E_1(M+1,S_z,\pm P)\,\,<\,\,E_2(M+1,S_z,\pm P)
\end{equation}
where the eigenenergies $E_i$ have been ordered from lower to higher
energy within each subspace.
See Table (cases A and B).
The difference in energies in the excited configurations come from the
decrease of Coulomb interaction due to the expansion of the
eigenstate of lower energy. The highest energy eigenstate
is a compact state and does not change in size. The two
excitations of case $A$ with and without Coulomb contribution are
equivalent
to the "sum mode" and the "difference mode" found by Girvin and
MacDonald \cite{gir} for DL (without tunneling or Zeeman terms and
for filling factor $\nu=1/2$) within a single-mode approximation.
Once we know the decrease of energy of Coulomb origin from
$E_2(10,7/2,0)-E_1(10,7/2,0)=E_c= 0.0998$, one can distinguish in the
$(11,7/2,0)$
configuration the double dipole $E_6=E_1(9,7/2,0)\,+\,2\alpha =
16.0802$ ($\alpha=0.2$)
from a quadrupole excitation given by
$E_4=E_1(9,7/2,0)\,+\,2\alpha\,-\,E_c = 15.9803$. Surprisingly, the
appearance of this eigenenergy in the output means that an eigenstate
which is not a compact state nor a fully expanded state, is a
possible realization. The full expansion would imply a larger value of
$E_c$ as
$\Delta M\,\,=\,\,2$.

\medskip

However, not all the previous scenario is reproduced in the $N=6$
system. If by analogy we consider the excitations
$(M,S_z,P)\,\rightarrow \,
(M+1,S_z,\pm P)$ from the $\mid$FM$\rangle$ GS, the energy
$E_2(M+1,S_z,P)\,\,=\,\,E_1(M,S_z,P)\,\,+\,\,\alpha$ is obtained
according to the behavior of the $N=7$ system but no trace of the
$\Delta_t$ energy ($\Delta_t=0.06$) is present in $E_2(M+1,S_z,-P)$.

\medskip

From the analysis of the results of the $N=7$ system, one would
suspect that the CM excitations (of energies $\alpha$ or
$\alpha\,+\,\Delta_t$) are a consequence of the Kohn theorem
in the DQD system, in agreement with our assumption of a parabolic
confining potential and
dipole-excitation approximation. However, if it would be the case, the
$N=6$ system would behave in the same way.
The reason for the difference is provided by the different action of
the electric fields that excite the system. The field in plane
produces a global shift
of the system which does not affect the internal electron-electron
interaction and the CM absorbs the total energy $\alpha$. In
contrast, if the external field has non-vanishing
component along the
$Z$-direction,
aside of the global absorbtions of energy $\alpha$ an extra amount of
energy is involved in a tunneling process.
In the odd electron system, the unpaired electron tunnels from one dot
to the other. This change, however, does not modify the electronic
distribution asymmetry which was already present in the GS and
consequently does not modify the Coulomb interaction. On the other
hand, in the even electron system due to the electronic jump, a change
is produced between
a symmetric distribution to a non-symmetric one giving necessarily
Coulomb contribution to the final energy.
That is to say, the excitations induced by $E_{\perp}$ on
an odd number
of electrons does not modify the Coulomb energy. However this is not a
consequence of the Kohn theorem, as one may erroneously conclude.

Within the second main point, the search of excitations that mark the
phase transitions, we concentrate on
the evolution of the system as
$\Delta_t$ increases, leaving all the other parameters unchanged. The
appropriate excitations turned
out to be the lowest energy  excitations.
Figs.(1) and (2)
show the excitation energy as a function of $\Delta_t$ for the
four different GS phases existing along the $\Delta_t$ energy
variation line for $N=6$ and $N=7$.
In all the cases the angular momentum is preserved and a simultaneous
spin  and  parity flip takes place. That is to say,

\begin{equation}
\omega \,\,=\,\,E_1(M,S_z\pm 1,-P)\,\,-\,\,E_1(M,S_z,P)
\end{equation}
where $E_1(M,S_z,P)$ is the GS energy at $\Delta_t$ and
$E_1(M,S_z
\pm
1,
-P)$ is the lowest eigenenergy within the
excited configuration at the same $\Delta_t$.
These excitations provide the lowest value ($\,\Delta M
=l\,=\,0\,$) of the multipole dispersion relation $\omega
(l)$ given below.

\medskip

Figs.(3) and (4) show the dispersion relation of the two
different types of multipole modes
$\omega (l)$ mentioned previously within the third main point. In both
cases, the lowest energy eigenstate within each subspace is
considered.
The values of
$\Delta_t$ were chosen in such a
way that for the two systems the GS is the first canted state: (6, 2,
1) for
N=6 and (9, 5/2, 1) for N=7.
The general trend is given by the dashed line
which corresponds to,

\begin{equation}
\omega\,\,=\,\,E_1(M+l,S_z,P)\,\,-\,\,E_1(M,S_z,P)
\end{equation}
The increase of angular momentum in one unit increases the kinetic
energy of the system by $\alpha$ ($\alpha=0.2$). This increase is
partially
compensated by the decrease of Coulomb energy due to expansion. For
both systems, the Coulomb contribution is numerically, for the
particular values of the parameters that we have taken, about
$2\alpha/3$ in such
a way that the total amount of energy gained in each step is about
$\alpha/3$. This gives a quite monotonous behavior.

\medskip

In contrast, unexpected features were obtained for excitations of the
type (solid line):

\begin{equation}
\omega\,\,=\,\,E_1(M+l,S_z+1,-P)-E_1(M,S_z,P)
\end{equation}
where spin and parity flip takes place simultaneously.
For low values of $l$, the contribution of the Coulomb energy to
$\omega$ is given by a more or less constant amount of, numerically,
$\alpha/2$.
However, after several steps, $4$ for N=6 and $6$ for N=7 the system
suffers a sudden expansion that reduces drastically the Coulomb
interaction. In the N=7 case, due to the presence of an unpaired
electron, the amount of the Coulomb contribution is so large that the total
energy decreases.
Looking at the occupancies of the single particle states of the
excited systems the nature of the sudden change becomes clear. For the
ferromagnetic excited states of the type $(6+l,\,\, 3, \,\,0)$ for
N=6 and
$(9+l,\,\, 7/2, \,\,0)$ for N=7 only spin-up states can exist. The
occupied single
particle states are $\mid$$s$$\uparrow\rangle$ and
$\mid$$a$$\uparrow\rangle$. As
$l$ increases from $l=0$, there
is a slow transfer of electrons from $\mid$$a$$\uparrow\rangle$ to
$\mid$$s$$\uparrow\rangle$.
The sudden
change takes place when the following structure is possible: N-1
electrons in $\mid$$s$$\uparrow\rangle$ with $\,m=0,..,N-2\,$ and one
electron in
$\mid$$a
$$\uparrow\rangle\,$
with m=0, which means M=10 (6+4) for N=6 and M=15 (9+6) for N=7. For
this eigenstate almost all the weight is in a single
Slater determinant which means that the system is nearly uncorrelated.

\medskip

The unexpected result is in clear contrast with the result obtained
for a single QD at the magic values of the angular
momentum, where all the energy is absorbed by the CM.
In the DQD most of the absorbed energy is
transformed into internal energy releasing the electrons from their
interaction.

\subsection*{IV Discussion and Conclusions}

We have analyzed several types of low energy excitations
over the three
possible GS of a DQD confined by parabolic potentials in each plane
and separated by a distance $d$. The LLL regime was considered and the
input parameters were
chosen in such a way that the filling factor of the ground states
and some of the excited states is close to
$\nu=2$.

\medskip

From the study of several dipole excitations ( $\Delta M\,=\,l\,=1$) we
verify that
the Kohn theorem is preserved in a DQD if the incident homogeneous
electric field is directed along the plane of the dots. For systems
with an odd number of electrons,
additional excitations independent of the Coulomb interaction, as the one
induced by $E_{\perp}$, are possible. This last possibility is due to
the fact that in this case, the isospin
change for a single electron does not change the Coulomb energy of the
system, although it is not a consequence of the Kohn theorem.

\medskip

The lowest energy excitations are characterized by a
simultaneous spin and
parity flip keeping the angular momentum unchanged (
spin-density-waves (SDW's) for $l\,=\,0$, with energy defined by
$\omega$  see Figs. 1 and 2). They turn out to be
the appropriate excitations which become degenerated
with the GS at
the transition boundaries in the
phase diagram.

\medskip

Due to the extra degree of freedom (as compared with a single dot or a
single layer), represented by the isospin states and due to the
interplay between tunneling and Coulomb interaction, the energy
$\omega$ has a Coulomb contribution. This is in contrast to
the case of a single layer for which, in the limit $k\,\rightarrow\,0$ the
non-interacting contribution given by $\Delta_z$ is recovered.

\medskip

The softening of the $\omega$ modes with the variation of
$\Delta_t$ shows phase instabilities previously detected in the
determination of the GS phase diagram from exact diagonalization
\cite{mar}.
There are some similarities and some differences between
our results and those reported by Das Sarma et al. \cite{das} within a
HF calculation for a double layer system. We will follow the
arguments given by these authors to analyze our results.
Within the
$\mid$FM$\rangle$ and $\mid$SYM$\rangle$ phases, the structure of the
curves $\omega/\Delta_t$ is similar, in both cases the
$\omega$ mode softens as it approaches the phase
transition from
$\mid$FM$\rangle$ to $\mid$C$\rangle$ and from $\mid$SYM$\rangle$ to
$\mid$C$\rangle$ respectively. In addition, the increase of
$\omega$ as it moves away from the boundary is larger in
the
symmetric phase in both calculations. However, within the canted state
we obtain finite values of $\omega$ although much smaller
than those in the $\mid$FM$\rangle$ or $\mid$SYM$\rangle$ phases, in
contrast to the results obtained in the DL for which
$\omega\,=\,0$ over the full canted phase. Aside from the
previous
outcomings, there is another main difference: the canted state in our
calculation is an eigenstate of the $S_z$ operator whereas this is not
the case for the HF canted phase. At the boundaries separating
different phases, however,
 due to energy
degeneracy, the superposition of states of different well defined spin
gives rise to states which are not eigenstates of $S_z$.
Das Sarma et al. \cite{das} have proved that the existence
of a gapless mode is directly due to the canted antiferromagnetic
spin ordering. As a consequence, even
though the canted states in
a DQD may be interpreted as having antiferromagnetic order in the
plain of the dots \cite{mar}, the order is not
complete, producing gapped excitations probably due to edge effects.
The exceptions are at the boundaries separating different phases where
due to degeneracy, the gapless mode is recovered.

\medskip

Finally, unexpected nearly uncorrelated states have been found (see
Figs. 3 and 4), produced by SDW excitations over the canted ground
states,
that develop a sudden expansion leaving the system in a ferromagnetic
state close to $\nu=1$. Besides that, the structure
developed
by the sudden expansion on the dispersion relation at a particular
value of $l$ is understood as a consequence of the existence of an
intrinsic
length scale in the system, related to the antiferromagnetic spin
ordering in plane. The structure is well defined in the SDW
dispersion relation (solid line in Figs. 3 and 4) and only
slightly apparent in the charge density wave curve (dashed line in
Figs.
3 and
4) due to the fact that within the LLL, local variation of spin is
coupled to variation of density \cite{moo} producing a larger
effect when spin and charge density fluctuations are induced
simultaneously by the excitation.
No structure was
found in similar dispersion relations of SDW over the ferromagnetic or
symmetric phases. In these last cases the dispersion is given by
nearly parabolic smooth curves whereas, over the canted states, in
the
long-wavelength limit, the dispersion relation is found to be linear,
consistent with the fact that it describes antiferromagnetic
fluctuations.

\medskip

Within the HF calculation performed by Das Sarma for DL, the SDW
dispersion relation does not have this type of structure. This difference
may be related to the different role that the kinetic contribution
to the energy plays in extended and in confined systems.

\medskip

Finally, in the limit $l\rightarrow \infty$ the dispersion relation of
the
SDW does not approach asymptotically a constant value as it is the case
for a single layer (SL)
or DL systems due to the fact that at this limit the excitation energy
always recovers the single particle value. For SL or DL this energy
is given by $\omega^{\prime}+V_{ex}$ where $\omega^{\prime}$ is the
non-interacting
excitation value ( a combination of $\Delta_z$ and $\Delta_t$) and
$V_{ex}$ is the exchange single particle energy of an electron in the
GS, independent of the linear momentum $k$ due to the degeneracy
existing in extended systems. In contrast, in a DQD the parabolic
potential breaks the degeneracy
producing an increase of energy with increasing angular momentum.
This gives a nearly linear curve as $l\rightarrow \infty$ typical
of a single particle.

\medskip

No total spin or space correlations in the ground states have been
investigated through the density-correlation functions which may signal
additional symmetry breaking
 effects. Symmetry breaking effects of this kind have been
found previously in literature for QD's, DQD's (vertical arrangement)
and
QD-molecules (horizontal arrangement) within unrestricted Hartree-Fock
calculations for low magnetic fields \cite{ya1,ya2} and within
Hamiltonian diagonalization for high magnetic fields \cite{ron}.
Whereas a comparison with the case of low magnetic fields is not
possible because we have strong magnetic fields which project the
system to the LLL, it would be interesting to see if our method, which
includes spin degrees of freedom and does not require any truncation,
shows symmetry breaking effects similar to Ref.\cite{ron}. This is
left for future investigations.

\bigskip 

{\bf Acknowledgements}

\medskip
 
We gratefully acknowledge C. Tejedor and L. Mart\'\i n-Moreno for helpful
discussions and for the code used for the Hamiltonian diagonalization.
This work has been performed under Grants No. BFM2002-01868,
 No. FPA2001-3598 from MCyT and Feder (Spain)
, and No. 2001SGR-0064 and No. 2001SGR-00065
from Generalitat de Catalunya .


\section{Appendix}

For simplicity we neglect spin indices since they are irrelevant for
the following discussion. We characterize the first quantized wave
function of the $N$ particle system as
$\Psi_{a_1,..., a_N}({\bf x_1},....,{\bf x_N})$, where $a_i=1,2$ are
layer indexes indicating the dot in which the i-th electron sits.
Hence the first quantized Hamiltonian is a matrix in layer-index
space
$H=H_{b_1 a_1;...;b_N a_N}$.

An in-plane homogeneous electric field clearly produces a  contribution
proportional to the identity matrix in the layer space,
$$
H_{E_{\Vert}}\sim \sum_{i=1}^{N}{\bf x_i}{\bf E_{\parallel}}
\left( \prod_{i=1}^{N}\delta_{b_i a_i}\right)
$$
and hence the interaction only depends on the center of mass coordinates,
namely the Kohn theorem applies. A homogeneous electric field
perpendicular
to the plane $E_{\perp}$ produces a contribution diagonal in layer
space
(but not proportional to the identity matrix) which reads
$$
H_{E_{\bot}}\sim\sum_{i=1}^{N} {d\over 2}(\delta_{a_i 1}-\delta_{a_i 2})
E_{\perp}
\left( \prod_{i=1}^{N}\delta_{b_i a_i}\right)
$$
where $d$ is the distance between the two dots.

The hopping term is not diagonal in layer space,
$$
H_{hop}\sim \Delta_t\sum_{i=1}^{N}\left( \prod_{j=1}^{i-1}\delta_{b_j
a_j}\right) s_{b_i a_i}\left( \prod_{j=i+1}^{N}\delta_{b_j a_j}\right)
$$
where $s_{11}=s_{22}=0$ and $s_{12}=s_{21}=1$. Then it is easy to check
that
$$
\left[ H_{E_{\bot}}, H_{hop} \right] \sim  \Delta_t{d\over 2}E_{\perp}
\sum_{i=1}^{N}
\left( \prod_{j=1}^{i-1}\delta_{b_j a_j}\right)
\epsilon_{b_i a_i}\left( \prod_{j=i+1}^{N}\delta_{b_j a_j}\right)
$$
where $\epsilon_{b_i a_i}$ is the antisymmetric tensor.
Hence a homogeneous electric field perpendicular to the plane changes
the dynamics in layer space in a non-trivial way. Since the latter
is entangled with the relative motion through the Coulomb term we
conclude that a homogeneous electric field perpendicular to the plane
may produce transitions between different states with the same center
of mass quantum numbers, namely the Kohn theorem does not apply.

\medskip
\medskip


\eject

\begin{figure}[htbp]
 \begin{center}
  \psfig{figure=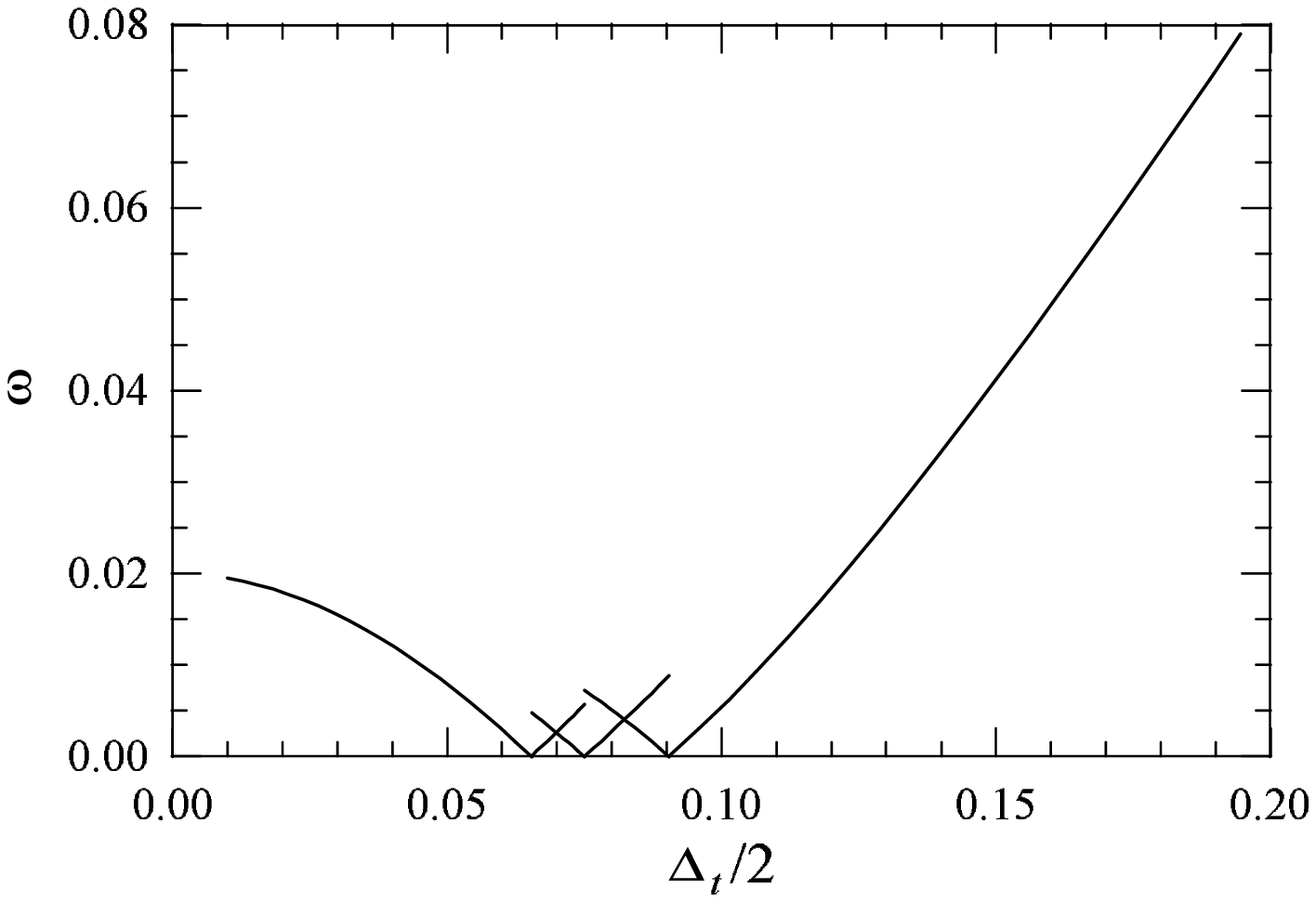,width=12.5cm}
 \end{center}
\caption{Lowest energy
excitation for $N=6$ as a function of $\Delta_t$ for different regions
of
the phase diagram. The excited configurations are (from left to right)
$(6,3,0)$,
$(6,2,1)$ and $(6,1,0)$  for the positive-gradient curves and
$(6,2,1)$,
$(6,1,0)$ and $(6,0,1)$ for the negative-gradient curves. The values
$\alpha=0.2$,
$\Delta_z=0.02$ and
$\beta=1.4$ have been considered.}
\end{figure}
\eject

\begin{figure}[htbp]
 \begin{center}
  \psfig{figure=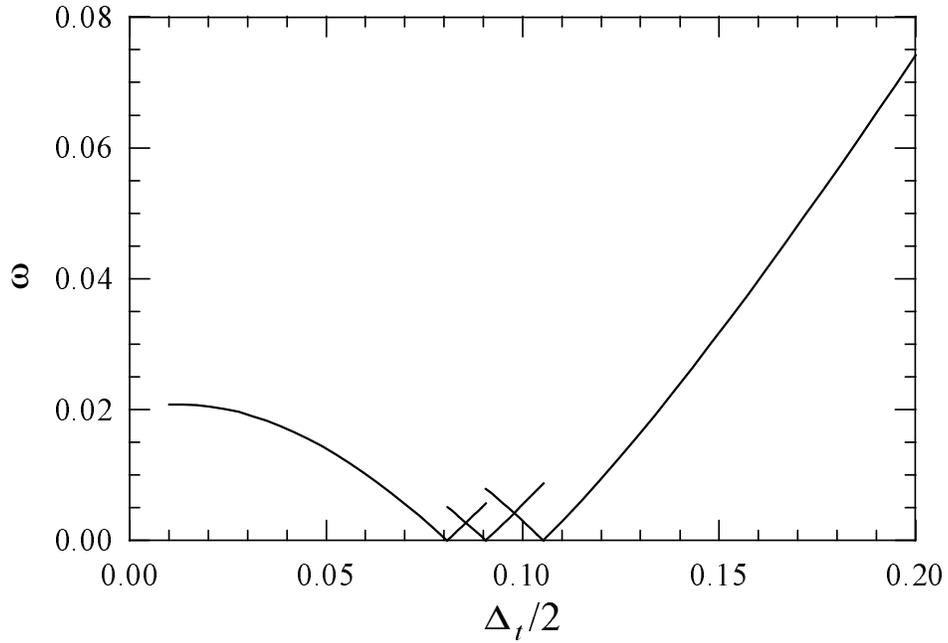,width=12.5cm}
 \end{center}
\caption{The same as Fig.1 for
$N=7$.
The excited configurations are (from left to right)
$(9,7/2,0)$,
$(9,5/2,1)$ and $(9,3/2,0)$  for the positive-gradient curves and
$(9,5/2,1)$,
$(9,3/2,0)$ and $(9,1/2,1)$ for the negative-gradient curves.}
\end{figure}
\eject

\begin{figure}[htbp]
 \begin{center}
  \psfig{figure=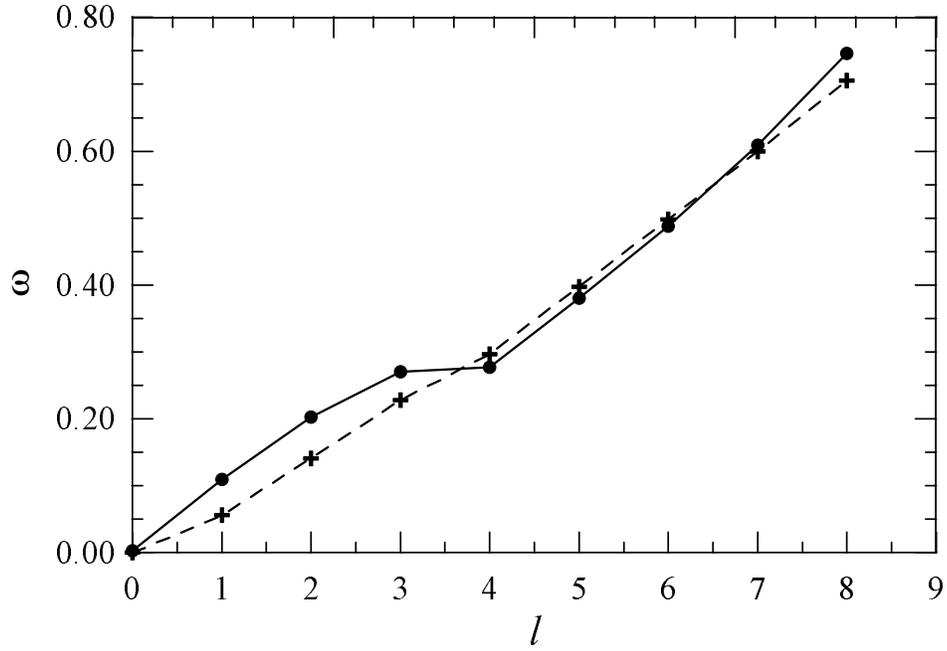,width=12.5cm}
 \end{center}
\caption{Dispersion relation $\omega (l)$ of two different
excited states: $(6+l,3,0)$ (solid line) and $(6+l,2,1)$ (dashed
line) for $N=6$ and
$\Delta_t=
0.07$. The GS is the first canted state.
The same values of $\alpha$,
$\beta$ and $\Delta_z$ as in Fig.1 have been used.}
\end{figure}
\eject

\begin{figure}[htbp]
 \begin{center}
  \psfig{figure=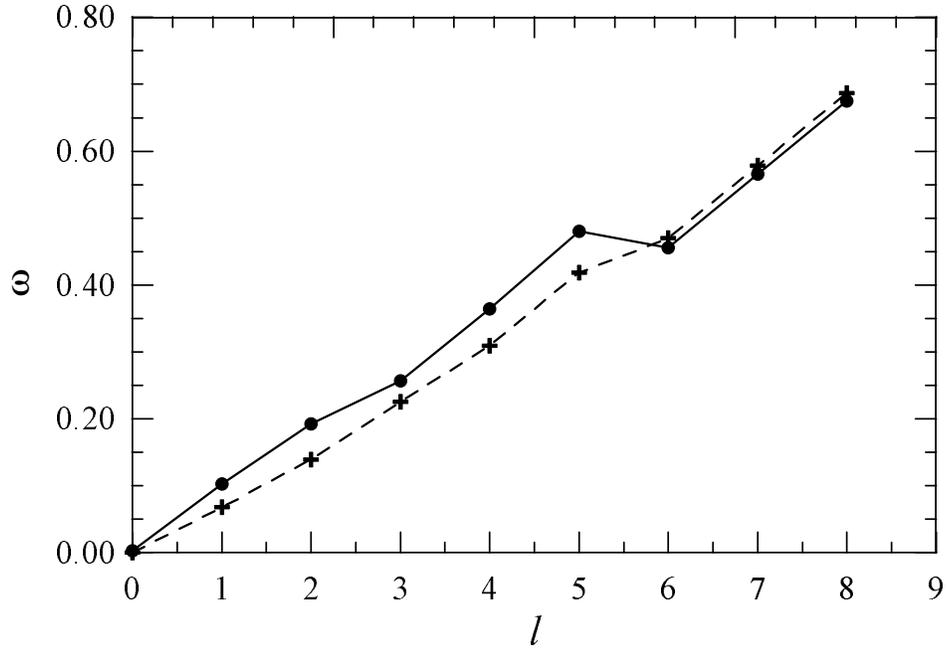,width=12.5cm}
 \end{center}
\caption{The same as in Fig.3 for the
excited states: $(9+l,7/2,0)$ (solid line) and $(9+l,5/2,1)$ (dashed
line) for $N=7$ and
$\Delta_t=
0.085$.}
\end{figure}
\eject

\begin{figure}[htbp]
 \begin{center}
  \psfig{figure=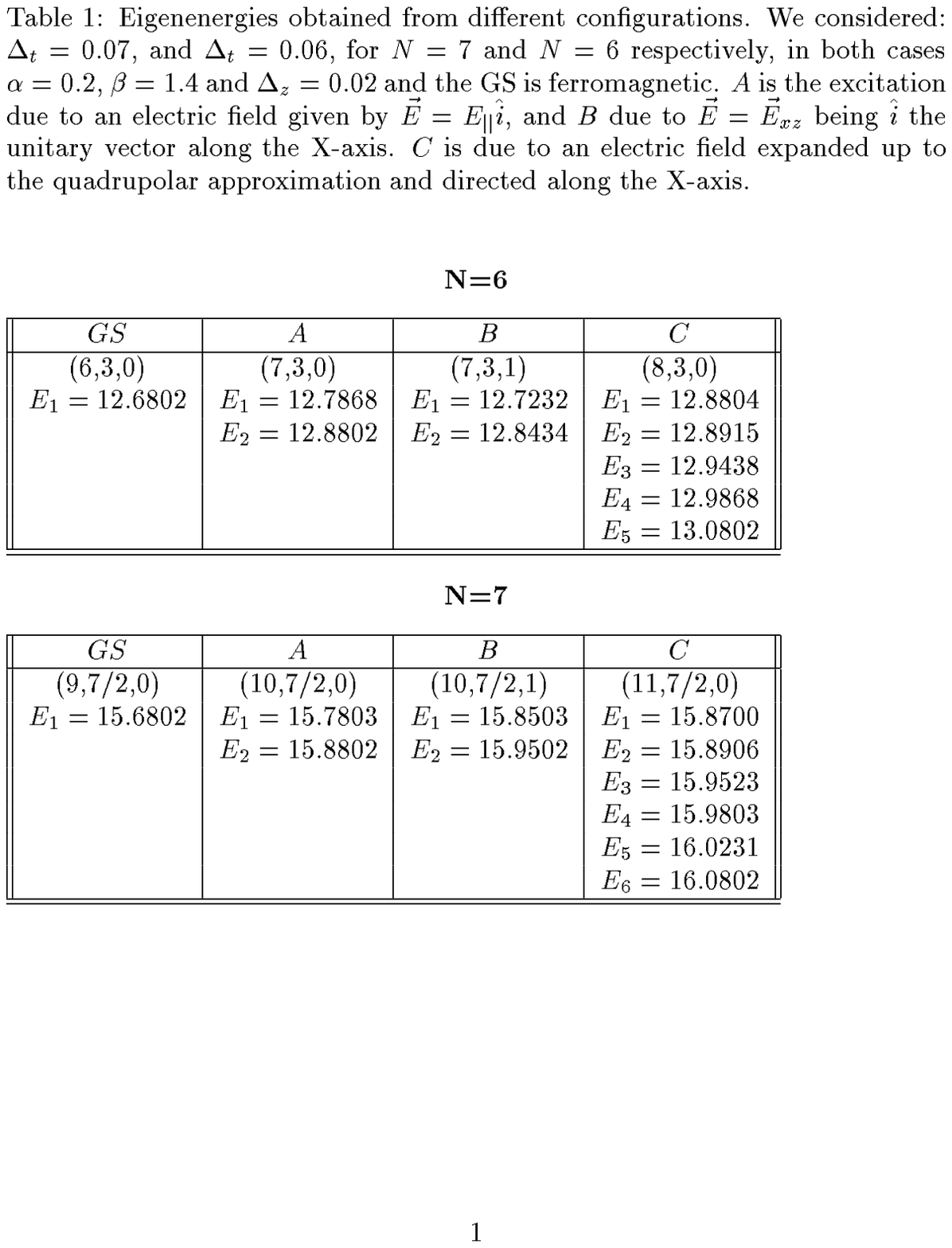,width=12.5cm}
 \end{center}
\end{figure}


\begin{references}

\bibitem{mar}
L. Martin-Moreno, L. Brey and C. Tejedor, Phys. Rev. ${\bf B62}$,
R10633 (2000).

\bibitem{par}
B. Partoens and F.M. Peeters, Phys. Rev. Lett. ${\bf 84}$, 4433
(2000).

\bibitem{tej}
C. Tejedor, B. Paredes, J.H. Oaknin, L. Martin-Moreno, Solid State
Comm. ${\bf 117}$, 133 (2001).

\bibitem{oak}
J.H. Oaknin, L. Martin-Moreno and C. Tejedor, Phys. Rev. ${\bf
B54}$, 16850 (1996).

\bibitem{moo}
K. Moon, H. Mori, Kun Yang, S.M. Girvin, A.H. MacDonald, L. Zheng, D.
Yoshioka, Shou-Cheng Zhang, Phys. Rev.
${\bf B51}$,
5138
(1995).

\bibitem{gir}
S.M. Girvin and A.H.MacDonald in "Perspectives in Quantum Hall
Effect", edited by Das Sarma and A. Pinczuk, Wiley N.Y. (1997).

\bibitem{das}
S. Das Sarma, S. Sachdev and L. Zheng, Phys. Rev. ${\bf B58}$,
4672 (1998).

\bibitem{son}
S.L. Sondhi, A. Karlhede, S.A. Kivelson and E.H. Rezayi, Phys. Rev.
${\bf B47}$, 16419 (1993).

\bibitem{par1}
B. Paredes, C. Tejedor, L. Brey and L. Martin-Moreno, Phys. Rev Lett
${\bf 83}$, 2250 (1999).

\bibitem{koh}
W. Kohn, Phys. Rev. ${\bf 123}$, 1242 (1961).

\bibitem{jac}
L. Jacak, P. Hawrylak and A. W¢js, "Quantum Dots" (Springer Verlag,
Berlin, 1998).

\bibitem{bre}
L. Brey, N.F. Johnson and B.J. Halperin, Phys. Rev ${\bf B40}$, 10647
(1989).

\bibitem{ya1}
C. Yannouleas and U. Landman, Phys. Rev. Lett. ${\bf
82}$, 5325 (1999).

\bibitem{ya2}
C. Yannouleas and U. Landman, Eur. Phys. J. ${\bf D16}$,
373 (2001).

\bibitem{ron}
M. Rontani, G. Goldoni, F. Manghi and E. Molinari, Europhys. Lett.
${\bf 58}$, 555 (2002).

\end{references}
\end{document}